\documentclass[pra,10pt,a4paper,twocolumn,english,noshowpacs,noshowkeys,notitlepage,superscriptaddress]{revtex4}

\usepackage{mathptmx}

\usepackage[T1]{fontenc}
\usepackage[latin9]{inputenc}
\usepackage{babel}
\usepackage[dvipdfm]{graphicx}

\begin{document}

\title{Generation of two-photon EPR and W states}

\author{W.B. Cardoso}
\affiliation{Instituto de F\'{\i}sica, Universidade Federal de Goi\'{a}s, 74.001-970, Goi\^{a}nia,
Goi\'{a}s, Brazil}
\author{W.C. Qiang}
\affiliation{Faculty of Science, Xi$^{\prime}$an University of Architecture and Technology, 710055, Xi$^{\prime}$an, China}
\author{A.T. Avelar}
\affiliation{Instituto de F\'{\i}sica, Universidade Federal de Goi\'{a}s, 74.001-970, Goi\^{a}nia,
Goi\'{a}s, Brazil}
\author{B. Baseia}
\affiliation{Instituto de F\'{\i}sica, Universidade Federal de Goi\'{a}s, 74.001-970, Goi\^{a}nia,
Goi\'{a}s, Brazil}

\begin{abstract}
In this paper we present a scheme for generation of two-photon EPR and W states in the cavity QED context. The scheme requires only one three-level Rydberg atom and two or three cavities. The atom is sent to interact with cavities previously prepared in vacuum states, via two-photon process. An appropriate choice of the interaction times allows one to obtain the mentioned states with maximized fidelities. These specific times and the values of success probability and fidelity are discussed.
\end{abstract}

\maketitle

The entanglement is present in a diversity of applications of quantum
mechanics, such as quantum teleportation \cite{BennettPRL93}, quantum
computation \cite{GottesmanNAT99}, one-way quantum computation \cite{RaussendorfPRL01},
quantum communication via teleportation \cite{CiracPRL97,BrassardPD98},
quantum metrology \cite{GilchristJOB04}, and quantum information
escaping from black holes \cite{LloydPRL06}. There are various kinds of 
entangled states depending on the number of involved parties and of the entanglement
`structure', e.g., for bipartite entanglements there is only one
example, the EPR state \cite{LoPRA01}; for 3-partite there are two kinds of
entangled states, the GHZ \cite{DurPRA00} and W \cite{AcinPRL01} states; for 4-partite,
more interesting is the 4-qubit cluster state whose correlations
cannot be described in terms of the local realism as
experimentally demonstrated in \cite{WaltherPRL05}. In fact, this state is not
biseparable and has a genuine 4-qubit entanglement.

With respect to the tripartite entangled states, it is known that
W states are robust against losses of qubits since they
retain bipartite entanglement if we trace out any one qubit,
whereas GHZ states are fragile since the remaining two
partite states result in separable states. This property
turns the W states very attractive for various quantum
information tasks, such as a quantum channel for teleportation
of entangled pairs \cite{GorbachevPLA03}, probabilistic teleportation \cite{JooNJP03}
and quantum key distribution \cite{Joo02}; however, one should manipulate it with care since it is not
appropriate for perfect quantum teleportation and superdense
coding in the form presented in Refs.\cite{KarlssonPRA98,PatiPRA00}. Several schemes have been proposed
to generate the W state in QED cavities \cite{GuoPRA02},
trapped ions \cite{LiPRA06}, quantum dots \cite{WangPRA03}, superconducting
quantum interference devices \cite{ZhengCPL07} and superconducting flux
qubits \cite{KimPRB08}.

A class of W states that can be used for perfect teleportation and superdense coding
was recently presented in \cite{AgrawalPRA06}, given by
\begin{equation}
|W_{\zeta}\rangle=\frac{1}{\sqrt{2+2\zeta}}(|001\rangle + \sqrt{\zeta}e^{i\gamma}|010\rangle 
+ \sqrt{\zeta+1}e^{i\delta}|100\rangle), \nonumber
\end{equation}
where $\zeta$ is a real number and $\gamma$, $\delta$
are phases. Additionally, in \cite{LiJPA07} Li and Qiu generalized the states of the W-class for
multi-qubit systems and multi-particle systems with higher dimension.

In view of applications of the entangled states discussed above, in this 
work we present a scheme for generation of two-photon EPR
and W states in QED cavities. To this end, a Rydberg three-level atom is sent to
interact with the cavities previously prepared in the vacuum state
via a two-photon process. It is worth to mention that the two-photon process has been
demonstrated in \cite{Brune87} for microwave cavity QED and offers
some advantages in relation to one-photon process, as the reduction
of interaction times due to the increasing of the atom-field
coupling strength and lower decoherence induced by stray fields
\cite{Blythe06}. In addition, two-photon process can be easily
obtained with Rydberg atoms with principal quantum number $n>89$,
which can be state-sensitively detected using tunneling field
ionization with quantum efficiencies above $80\%$ and an ionization efficiency above $%
98\%$ \cite{Tada02}.  In the following, we discuss the theoretical model and give details of our scheme.

\begin{figure}[b]
\centering
\includegraphics[width=3cm]{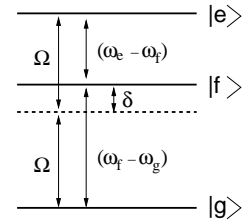}
\caption{Schematic diagram of the three-level atom interacting with a single-mode
of a cavity field.}
\label{levels}
\end{figure}

Consider a three-level atom that interacts with a single cavity-field mode via two-photon Jaynes-Cummings model described in the interaction picture by the Hamiltonian \cite{ToorPRA92}
\begin{eqnarray}
H_{I} &=&\hbar g_{1}\left( a|e\rangle \langle f|e^{-i\delta t}+a^{\dagger}
|f\rangle \langle e|e^{i\delta t}\right)   \nonumber \\
&&+\hbar g_{2}\left( a|f\rangle \langle g|e^{i\delta t}+a^{\dagger
}|g\rangle \langle f|e^{-i\delta t}\right) ,  \label{2JCM}
\end{eqnarray}%
where $g_{1}$ and $g_{2}$ stand for the one-photon coupling constant with
respect to the transitions $|e\rangle \leftrightarrow |f\rangle $ and $%
|f\rangle \leftrightarrow |g\rangle $, respectively. The detuning $\delta $
is given by
\begin{equation}
\delta =\Omega -(\omega _{e}-\omega _{f})=(\omega _{f}-\omega _{g})-\Omega ,
\label{eq:1}
\end{equation}%
where $\Omega $ is the cavity-field frequency and $\omega _{e}$, $\omega _{f}
$, and $\omega _{g}$ are the frequencies associated with the atomic levels $%
|e\rangle $, $|f\rangle $, and $|g\rangle $, respectively. Fig. \ref{levels} shows a
schematic representation of the atomic levels.

The state describing the combined atom-field system reads
\begin{equation}
|\psi (t)\rangle =\sum_{n}\left[ C_{e,n}(t)|e,n\rangle
+C_{f,n}(t)|f,n\rangle +C_{g,n}(t)|g,n\rangle \right] ,  \label{eq:4}
\end{equation}%
where $|k,n\rangle ,$ with $k=e,$ $f,$ $g,$ indicating the atom in the state $%
|k\rangle $ and the field in the Fock state $|n\rangle $. The coefficients $C_{k,n}(t)$
stand for the corresponding probability amplitudes.

Inserting the Eqs. (\ref{2JCM}) and (\ref{eq:4}) into the time dependent Schr%
\"{o}dinger equation one obtains the coupled first-order differential
equations for the probability amplitudes
\begin{eqnarray}
\frac{dC_{e,n}(t)}{dt} &=&-ig_{1}C_{f,n+1}(t)\sqrt{n+1}e^{-i\delta t},
\nonumber \\
\frac{dC_{f,n+1}(t)}{dt} &=&-ig_{1}C_{e,n}(t)\sqrt{n+1}e^{i\delta t}  \nonumber
\\
&&-ig_{2}C_{g,n+2}(t)\sqrt{n+2}e^{i\delta t},  \nonumber \\
\frac{dC_{g,n+2}(t)}{dt} &=&-ig_{2}C_{f,n+1}(t)\sqrt{n+2}e^{-i\delta t}.
\label{eq:5}
\end{eqnarray}

As usually, we consider that the entire atom-field system is decoupled at
the initial time $t=0,$%
\begin{eqnarray}
C_{e,n}(0) &=&C_{e}C_{n}(0),  \nonumber \\
C_{b,n+1}(0) &=&C_{f}C_{n+1}(0),  \nonumber \\
C_{c,n+2}(0) &=&C_{g}C_{n+2}(0),  \label{eq:8}
\end{eqnarray}%
where the $C_{n}(0)$ stand for the amplitudes of the arbitrary initial field
state and the $C_{a}$ are atomic amplitudes of the (normalized) initial
atomic state
\begin{equation}
|\chi \rangle =C_{e}|e\rangle +C_{f}|f\rangle +C_{g}|g\rangle .  \label{eq:7}
\end{equation}

\begin{widetext}
Solving these coupled differential equations with the initial conditions in (\ref{eq:8})
we get the time dependent coefficients as
\begin{eqnarray}
C_{e,n}(t)&=&\left[ \frac{g_{1}^{2}(n+1)}{\Lambda _{n}\alpha _{n}^{2}}\gamma
_{n}(t)+1\right] C_{e}C_{n}-i\frac{g_{1}\sqrt{n+1}}{\Lambda _{n}}\sin
(\Lambda _{n}t)e^{-i\frac{\delta t}{2}}C_{f}C_{n+1} \nonumber \\
&&+\left[ \frac{g_{1}g_{2}%
\sqrt{(n+1)(n+2)}}{\Lambda _{n}\alpha _{n}^{2}}\gamma _{n}(t)\right]
C_{g}C_{n+2},  \label{eq:9}
\end{eqnarray}%
\begin{eqnarray}
C_{f,n+1}(t)&=&-i\frac{g_{1}\sqrt{n+1}}{\Lambda _{n}}\sin (\Lambda _{n}t)e^{i%
\frac{\delta t}{2}}C_{e}C_{n}+\left( \cos (\Lambda _{n}t)-\frac{i\delta }{%
2\Lambda _{n}}\sin (\Lambda _{n}t)\right) e^{i\frac{\delta t}{2}%
}C_{f}C_{n+1} \nonumber \\
&&-i\frac{g_{2}\sqrt{n+2}}{\Lambda _{n}}\sin (\Lambda _{n}t)e^{i%
\frac{\delta t}{2}}C_{g}C_{n+2},  \label{eq:10}
\end{eqnarray}%
\begin{eqnarray}
C_{g,n+2}(t)&=&\frac{g_{1}g_{2}\sqrt{(n+1)(n+2)}}{\Lambda _{n}\alpha _{n}^{2}}%
\gamma _{n}(t)C_{e}C_{n}-i\frac{g_{2}\sqrt{n+2}}{\Lambda _{n}}\sin (\Lambda
_{n}t)e^{-i\frac{\delta t}{2}}C_{f}C_{n+1} \nonumber \\
&&+\left[ \frac{g_{2}^{2}(n+2)}{%
\Lambda _{n}\alpha _{n}^{2}}\gamma _{n}(t)+1\right] C_{g}C_{n+2},
\label{eq:11}
\end{eqnarray}
\end{widetext}%
where
\begin{equation}
\gamma _{n}(t)=\left[ \Lambda _{n}\cos (\Lambda _{n}t)+i\frac{\delta }{2}%
\sin (\Lambda _{n}t)-\Lambda _{n}e^{i\frac{\delta t}{2}}\right] e^{-i\frac{%
\delta t}{2}},  \label{eq:12}
\end{equation}%
\begin{equation}
\Lambda _{n}=\sqrt{\frac{\delta ^{2}}{4}+\alpha _{n}^{2}}~,  \label{eq:13}
\end{equation}%
\begin{equation}
\alpha _{n}=\sqrt{g_{1}^{2}(n+1)+g_{2}^{2}(n+2)},  \label{eq:14}
\end{equation}%
$\Lambda _{n}$ being the Rabi frequency. The substitutions $n\rightarrow n-1$
in Eq. (\ref{eq:10}) and $n\rightarrow n-2$ in Eq. (\ref{eq:11}) allow one
to obtain the $C_{f,n}(t)$ and $C_{g,n}(t)$, respectively.

\textit{Two-photon EPR state} - Now, we describe the scheme for generation of the maximally entangled two-photon
EPR state, written in the form%
\begin{equation}
|\psi\rangle_{12} = \frac{1}{\sqrt{2}} (|02\rangle_{12}+|20\rangle_{12}),
\end{equation}
where the subscripts 1 and 2 concern the cavities 1 and 2, respectively.
Assume the two cavities previously prepared in the vacuum state and the atom in its excited state ($|e\rangle$).

Firstly, the atom crosses the cavity 1, interacting with the field mode for a time
$t_1$. Thus, the initial state describing the atom plus the two cavities evolves to%
\begin{eqnarray}
|\phi\rangle_{a12} &=& C_{e0}^{(e0)}(t_1) |e00\rangle_{a12} +
C_{f1}^{(e0)}(t_1) |f10\rangle_{a12} \nonumber \\
&&+ C_{g2}^{(e0)}(t_1) |g20\rangle_{a12},
\end{eqnarray}
where $C_{ij}^{(k,l)}$ are the coefficients in Eqs. (\ref{eq:9})-(\ref{eq:11}) considering the input
atomic state in $k$ and the field state in $l$.

Next, the atom enters the cavity 2 and interacts for a time $t_2$. So,
the state of the whole system is written as%
\begin{eqnarray}
|\phi^{\prime}\rangle_{a12} &=& C_{e0}^{(e0)}(t_1)[ C_{e0}^{(e0)}(t_2) |e00\rangle_{a12} +
C_{f1}^{(e0)}(t_2) |f01\rangle_{a12} \nonumber \\
&&+ C_{g2}^{(e0)}(t_2) |g02\rangle_{a12}] +
C_{f1}^{(e0)}(t_1)[ C_{f0}^{(f0)}(t_2) |f10\rangle_{a12} \nonumber \\
&&+ C_{g1}^{(f0)}(t_2) |g11\rangle_{a12}]
+ C_{g2}^{(e0)}(t_1) |g20\rangle_{a12}.
\end{eqnarray}

\begin{figure}[t]
\includegraphics[width=4cm]{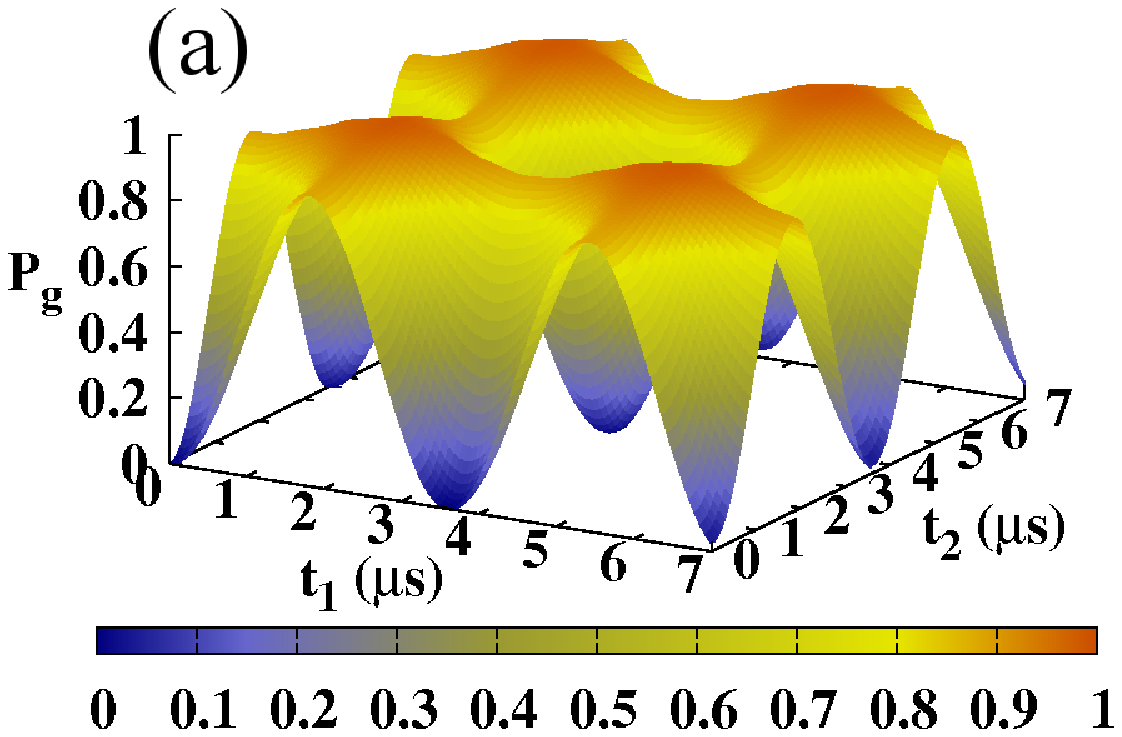}\hfil%
\includegraphics[width=4cm]{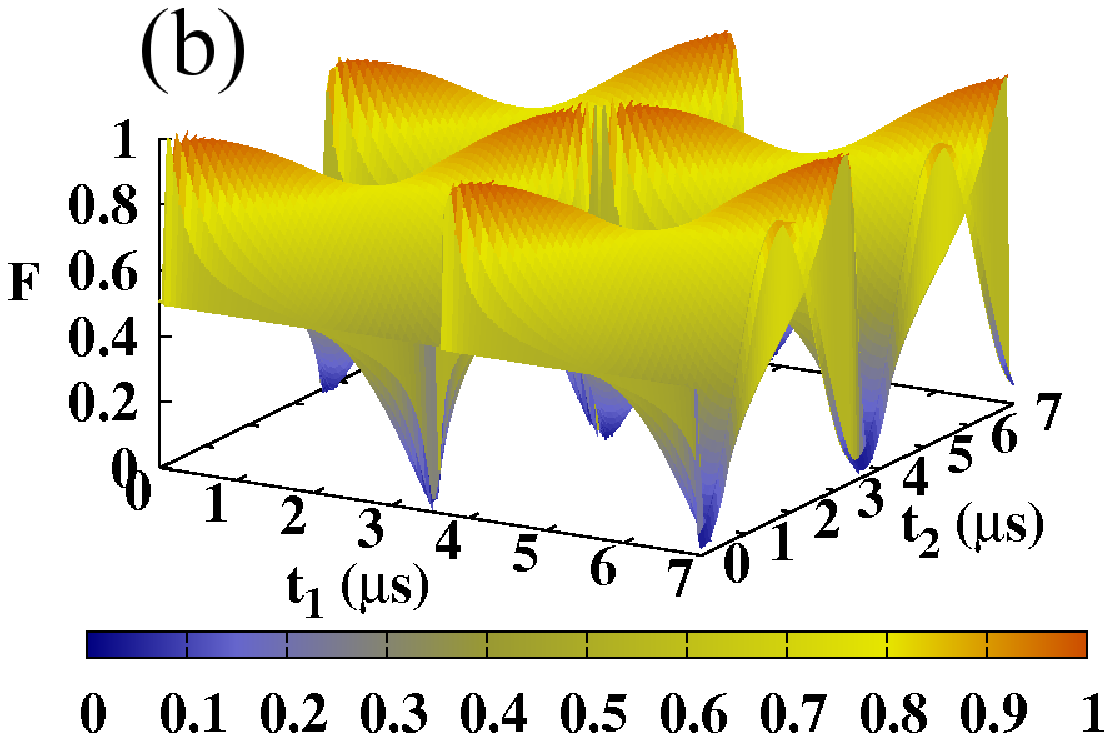}
\caption{(Color online) Plots of (a) success probability of detection of the 
atomic state in $|g\rangle$ and (b) the respective fidelity for the
two-photon EPR state, for parameters $g_1=g_2=g=17.5$MHz, 
$\delta=30g$.} \label{F1}
\end{figure}

\begin{figure}[t]
\includegraphics[width=4cm]{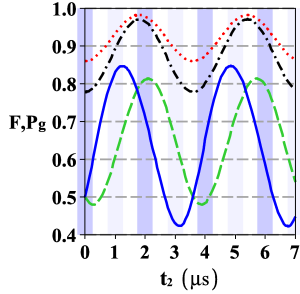}
\caption{(Color online) Comparison of the fidelity and success probability versus the time $t_2$. 
We use the values $t_1=2\mu s$ for the fidelity in dashed line (green) and for success
probability in dotted line (red) and $t_1=5\mu s$ for the fidelity in solid line (blue) 
and for the success probability in dash-dot line (black), with same conventions of Fig \ref{F1}.} %
\label{F2}
\end{figure}

To get the two-photon EPR state we may choose to realize the atomic detection or not. In the affirmative case,
the detection of the atomic state $|g\rangle$ leads to the success probability
\begin{eqnarray}
P_g &=& |C_{e0}^{(e0)}(t_1)|^2|C_{g2}^{(e0)}(t_2)|^2 + |C_{f1}^{(e0)}(t_1)|^2|C_{g1}^{(f0)}(t_2)|^2 \nonumber \\
&+& |C_{g2}^{(e0)}(t_1)|^2,
\end{eqnarray}
and the respective fidelity
\begin{eqnarray}
F &=& \frac{1}{2P_g}|C_{e0}^{(e0)}(t_1)C_{g2}^{(e0)}(t_2) + C_{g2}^{(e0)}(t_1)|^2.
\end{eqnarray}
The success probability and fidelity of the EPR state with atomic detection
is displayed in Fig. \ref{F1}a and \ref{F1}b, respectively. The values of experimental
parameters $g_1=g_2=g=17.5$MHz and $\delta=30g$ were used. In this case, Fig. \ref{F2} presents the values of the success probability and fidelity for two different times. Note that  
an appropriate choice of the interaction time allows one to obtain a greater fidelity.

\begin{figure}[t]
\includegraphics[width=6cm]{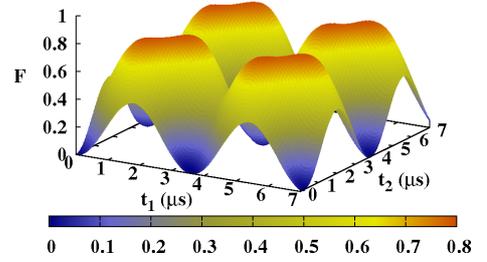}
\caption{(Color online) Fidelity of the two-photon EPR state without atomic detection, using the same conventions of Fig \ref{F1}.} %
\label{F3}
\end{figure}

On the other hand, if we choose to realize no atomic detection, the maximum value of the fidelity decreases
to a value close to $0.8$, with the form
\begin{eqnarray}
F &=& \frac{1}{2}|C_{e0}^{(e0)}(t_1)C_{g2}^{(e0)}(t_2) + C_{g2}^{(e0)}(t_1)|^2,
\end{eqnarray}
as used in the Fig. \ref{F3}.

\textit{Two-photon W state} - The W state to be constructed is written as
\begin{equation}
|\psi\rangle_{123} = \frac{1}{2}( |002\rangle_{123} + |020\rangle_{123} + \sqrt{2}|200\rangle_{123} ),
\end{equation}
where the subscripts concern the cavities 1, 2, and 3, respectively. Also, we can
generate the two-photon W state using a similar procedure to the
two-photon EPR state generation. However, the W state is realized in three cavities and a distinct procedure is required.

Firstly, we consider these three cavities in the vacuum state. An atom in the excited
state $|e\rangle$ is sent to interact with the first cavity by a time $t_1$, leading the
whole system in the state%
\begin{eqnarray}
|\phi\rangle_{a123} &=& C_{e0}^{(e0)}(t_1) |e000\rangle_{a123} +
C_{f1}^{(e0)}(t_1) |f100\rangle_{a123} \nonumber \\
&&+ C_{g2}^{(e0)}(t_1) |g200\rangle_{a123}.
\end{eqnarray}

Next, the atom crosses the cavities 2 and 3 by a time $t_2$ and $t_3$, respectively. The above state goes to the form
\begin{eqnarray}
|\phi^{\prime\prime}\rangle_{a123} &=& C_{e0}^{(e0)}(t_1)\{C_{e0}^{(e0)}(t_2)[
C_{e0}^{(e0)}(t_3)|e000\rangle_{a123} \nonumber \\
&&+ C_{f1}^{(e0)}(t_3)|f001\rangle_{a123} +
C_{g2}^{(e0)}(t_3)|g002\rangle_{a123} ] \nonumber \\
&&+ C_{f1}^{(e0)}(t_2)[ C_{f0}^{(f0)}(t_3)|f010\rangle_{a123} \nonumber \\
&&+ C_{g1}^{(f0)}(t_3)|g011\rangle_{a123} ] + C_{g2}^{(e0)}(t_2)|g020\rangle_{a123} \} \nonumber \\
&&+ C_{f1}^{(e0)}(t_1)\{ C_{f0}^{(f0)}(t_2) [ C_{f0}^{(f0)}(t_3)|f100\rangle_{a123} \nonumber \\
&&+ C_{g1}^{(f0)}(t_3)|g101\rangle_{a123} ]
+ C_{g1}^{(f0)}(t_2)|g110\rangle_{a123} \} \nonumber \\
&&+ C_{g2}^{(e0)}(t_1)|g200\rangle_{a123}.
\label{Eq19}
\end{eqnarray}

\begin{figure}[t!]
\includegraphics[width=4cm]{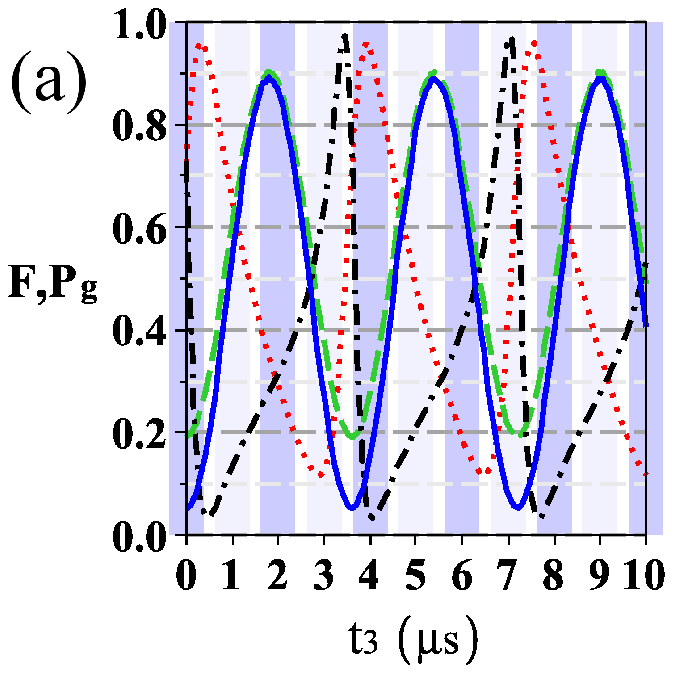}\hfil%
\includegraphics[width=4cm]{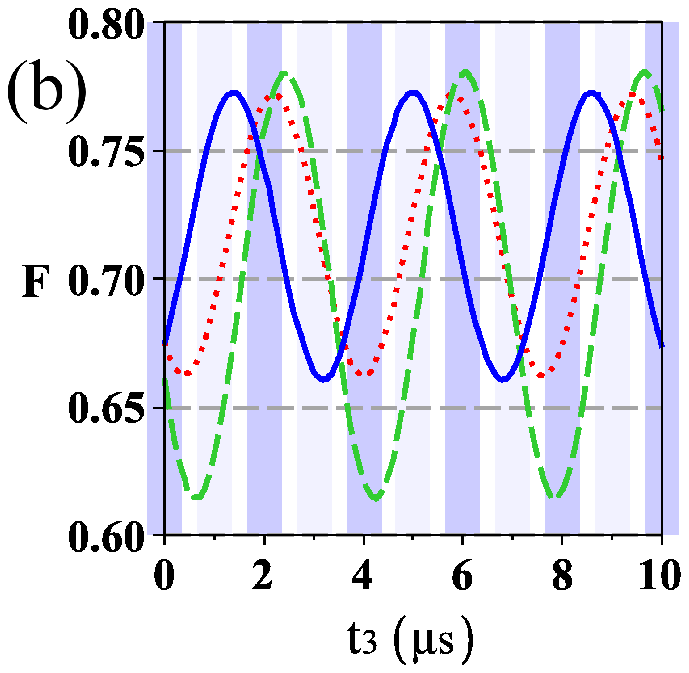}
\caption{(Color online) Plots of fidelity and success probability versus the interaction time $t_3$ 
for the two-photon W state generation scheme. In (a) we consider the atomic 
detection in the state $|g\rangle$, with the values $t_1=t_2=4\mu s$ for the fidelity 
in dotted line (red) and for success probability in dashed line (green) and the values 
$t_1=t_2=7\mu s$ for the fidelity in dash-dot line (black) and for success probability 
in solid line (blue). In (b) we consider the scheme without atomic detection. For this case 
we show three values of the fidelity versus $t_3$. In solid line (blue) we use $t_1=16\mu s$ and 
$t_2=16 \mu s$; in dotted line (red) we use $t_1=t_2=2\mu s$; in dashed line (green) we use 
$t_1=t_2=13\mu s$, using the same conventions of Fig \ref{F1}.} %
\label{F4}
\end{figure}

If we detect the atom in its ground state ($|g\rangle$) the state of the system 
collapses in the form
\begin{eqnarray}
|\phi^{\prime\prime\prime}\rangle_{123} &=& N \{ C_{e0}^{(e0)}(t_1)[C_{e0}^{(e0)}(t_2)
C_{g2}^{(e0)}(t_3)|002\rangle_{123} \nonumber \\
&&+ C_{f1}^{(e0)}(t_2) C_{g1}^{(f0)}(t_3)|011\rangle_{123}
+ C_{g2}^{(e0)}(t_2)|020\rangle_{123} ] \nonumber \\
&&+ C_{f1}^{(e0)}(t_1)[ C_{f0}^{(f0)}(t_2) C_{g1}^{(f0)}(t_3)|101\rangle_{123} \nonumber \\
&&+ C_{g1}^{(f0)}(t_2)|110\rangle_{123} ] + C_{g2}^{(e0)}(t_1)|200\rangle_{123},
\end{eqnarray}
with success probability given by

\begin{eqnarray}
P_g &=& |C_{e0}^{(e0)}(t_1)|^2|C_{e0}^{(e0)}(t_2)|^2|C_{g2}^{(e0)}(t_3)|^2 \nonumber \\
&&+ |C_{e0}^{(e0)}(t_1)|^2|C_{f1}^{(e0)}(t_2)|^2|C_{g1}^{(f0)}(t_3)|^2 \nonumber \\
&&+ |C_{e0}^{(e0)}(t_1)|^2|C_{g2}^{(e0)}(t_2)|^2 \nonumber \\
&&+ |C_{f1}^{(e0)}(t_1)|^2|C_{f0}^{(f0)}(t_2)|^2|C_{g1}^{(f0)}(t_3)|^2 \nonumber \\
&&+ |C_{f1}^{(e0)}(t_1)|^2|C_{g1}^{(f0)}(t_2)|^2 + |C_{g2}^{(e0)}(t_1)|^2.
\end{eqnarray}

We calculate the fidelity $F=|_{123}\langle\psi|\phi^{\prime\prime\prime}\rangle_{123}|^2$,
written as

\begin{eqnarray}
F &=& \frac{1}{4Pg} | C_{e0}^{(e0)}(t_1)C_{e0}^{(e0)}(t_2)C_{g2}^{(e0)}(t_3) \nonumber \\
&&+ C_{e0}^{(e0)}(t_1)C_{g2}^{(e0)}(t_2) + \sqrt{2} C_{g2}^{(e0)}(t_1) |^2.
\end{eqnarray}
With an appropriate choice of $t_i$ ($i=1,2,3$) one gets the maximized fidelity.
For example, in the case of atomic detections we choose the interaction times 
$t_1=t_2=t_3=32\mu$s to obtain approximately success 
probability $30\%$ and fidelity $95\%$.

In conclusion, we presented a scheme for the generation of two-photon EPR and W states in the cavity QED context. Concerning with the experimental feasibility, the devices used here were based on Ref. \cite{Blythe06} 
concerned with Rydberg atoms with quantum number $n\sim90$. The coupling constant 
of these atoms with the cavity is taken close to 	$g = 17.5$MHz. In our simulations we have used
this value and a detuning of $30g$. As example, by choosing the interaction times as $t_1=t_2=3\mu$s for the two-photon EPR state the success probability and fidelity result approximately $40\%$ and $97\%$, respectively; if we choose $t_1=t_2 = t_3=32\mu$s for the two-photon W state, the success probability and fidelity result $30\%$ and $95\%$, respectively.
On the other hand, the time spent for the entire procedure
is approximately $10^{-4}$s for both, EPR and W case. These times are much smaller than the decoherence time of the cavity ($0.1$s) \cite{GleyzesNAT05}, which shows the experimental feasibility of the scheme.

We thank the CAPES, CNPq, and FUNAPE/GO, Brazilian
agencies, for the partial supports. This work is also partially supported by
Natural Science Basic Research Plan in the Shaanxi Province of
China (program no: SJ08A13), the Natural Science Foundation of the
Education Bureau of Shaanxi Province, China under Grant O9jk534.

\end{document}